# New evidence for the fluctuation characteristic of intradecadal periodic signals in length-of-day variation

Hao Ding[*], Yachong An, Wenbin Shen

*Department of Geophysics, School of Geodesy and Geomatics, Key Laboratory of Geospace Environment and Geodesy of the Ministry of Education, Wuhan University, 430079, Wuhan, China*

**Corresponding address**: dhaosgg@sgg.whu.edu.cn

**Keypoints**

1. We prove that the ~5.9yr and ~8.5yr signals in the ΔLOD do not have stable damping trends but have time-varying amplitudes in time domain.

2. A possible 7.6yr periodic signal was found in the ΔLOD for the first time, we confirm that it is also present in the global geomagnetic records.

3. Both the ~5.9yr and ~8.5yr signals seem to be related to jerks, which suggest that jerks may be excitation sources for these oscillations.

**Abstract**

The intradecadal fluctuations in the length-of-day variation (ΔLOD) are considered likely to play an important role in core motions. Two intradecadal oscillations, with ~5.9yr and ~8.5yr periods (referred to as SYO and EYO, respectively), have been detected in previous studies. However, whether the SYO and the EYO have stable damping trends since 1962 and whether geomagnetic jerks are possible excitation sources for the SYO/EYO are still debated. In this study, based on different methods and ΔLOD records with different time span, we show robust evidences to prove that the SYO and the EYO have no stable damping trends since 1962, and we find that there is also a possible ~7.6yr signal. To prove whether it is a periodic signal, we use the optimal sequence estimation method to stack 35 global geomagnetic records, the results also show an ~7.6yr periodic signal which has an $Y_{2,-2}$ spatial distribution, and it has a high degree of consistent synchronicity with the ~7.6yr signal in ΔLOD. After confirming that the jerks have no special consistency with the peaks/valleys of the EYO/SYO, we confirm that the geomagnetic jerks seem to be related to sudden changes in the SYO/EYO time series and their excitation series; so we finally suggest that jerks are possible excitation sources of the SYO/EYO. Meanwhile, after using a deconvolution method, we estimate that the period $P$ and quality factor $Q$ of the SYO and the EYO are [$P$=5.85±0.06yr, $Q$≥180] and [$P$=8.45±0.17yr, $Q$≥350], respectively.

**Plain Language Summary**

An intradecadal periodic signal in the length-of-day variation (ΔLOD), the ~5.9yr oscillation (SYO), has been studied for decades; another ~8.5yr oscillation (EYO) has been detected and studied recently (Ding, 2019; Duan & Huang, 2020a). However, different researchers have determined different fluctuation characteristic, and whether the oscillations in the 1962-2019 time span are stable damping oscillations is still in dispute (e.g., Holme & de Viron, 2013 *Nature*; Duan et al., 2018,

*Earth Planet Sc. Lett.*; Ding, 2019, *Earth Planet Sc. Lett.*; Duan & Huang, 2020a, *Nature Commun.* and 2020b, *JGR-Solid Earth*). In addition, whether the SYO and the EYO have relationships with geomagnetic jerks is still in dispute. In this study, we mainly try to resolve these two disputes. Our results confirm that there is no damping trend for the SYO and the EYO, and we explain why the same damping results were obtained by previous results. Although our results also show that the SYO/EYO may have some relationships with jerks, different with the result and suggestion of a previous study, our results show that there is no particular consistency between the jerks and the peaks/valleys of the EYO, and the EYO seems to offer no remarkable help in predicting jerks.

**Keywords**

Intradecadal fluctuations; Length-of-day variations; Damping trends; Geomagnetic jerks

**1. Introduction**

The fluctuation characteristics and excitations of the intradecadal changes in the length-of-day variation (ΔLOD) were thought to be related to the short-period secular variations in the core geomagnetic field, and hence will help to constrain the strength of the magnetic field in the core and to understand the mechanism driving the Earth's core-mantle interactions (e.g., Mound & Buffett, 2006; Gillet et al., 2010; Holme & de Viron, 2013; Gross, 2015). Two periodic signals have been detected from the ΔLOD in the intradecadal period band (here we mean the 5-10yr period band), an approximately six year oscillation (SYO) (e.g., Liao and Greiner-Mai, 1999; Abarca del Rio et al.,

2000; Mound and Buffett, 2006; Holme and de Viron, 2013; Chao et al., 2014; Duan et al., 2015, 2017, 2018; Ding and Chao 2018a,b; Ding 2019; Duan and Huang, 2020a) and an approximately 8.5yr oscillation (EYO) (Ding 2019; Duan and Huang, 2020a). The fluctuation characteristics of those intradecadal changes are thought to help determine their possible mechanism (see Gillet et al., 2010, 2015, 2019). However, the fluctuation characteristics of these two signals are still controversial.

Liao and Greiner-Mai (1999) first found a nearly stable ~5.8yr oscillation in the 1970-1990 ΔLOD (see their Fig. 5a) and found that the Southern Oscillation Index may have some correlation with it. Abarca del Rio et al. (2000) also showed that the 6-7yr oscillation has no stable decreasing trend in the 1900-2000 time-span (see their Figs. 3 and 5), and its fluctuation characteristic is similar to a modulation, as suggested by Ding (2019; referred to as D19). Holme and de Viron (2013) showed that the SYO is a stable fluctuation in the 1962-2012 time span after using an iterative fitting and removal process (see their Fig. 2). Chao et al. (2014) showed the Morlet wavelet spectrum of the ΔLOD in the 1962-2012 time span, and their results roughly indicated that the SYO had a decreasing trend from 1965-1997, but changed to an increasing trend after 1997 (see their Fig. 1c; also similar as the finding in D19). After using a Daubechies wavelet low-pass filter and combined with a symmetric extension, Duan et al. (2015) used the normal Morlet wavelet (NMWT) method (Liu et al., 2007) for the extended ΔLOD time series and found that the SYO has a nearly stable decreasing trend in the 1962-2012 time span (see their Fig. 9). Based on the same method and same record (1962-2012), Duan et al. (2017) further proposed a damping model and estimated that the quality factor $Q$ of the SYO is 51.6±0.4 based on fitting the envelope curve of the SYO in the time domain.

Based on this $Q$ value and a free decay trend for the SYO, Duan et al. (2018) and Duan and Huang (2020b) further considered electromagnetic (EM) coupling at the core-mantle boundary (CMB) under the MICG mechanism. Based on the optimal sequence estimation method (Ding and Shen, 2013; Ding and Chao, 2015a), Ding and Chao (2018a) first found the SYO in global GPS and geomagnetic records, and their results showed that the SYO in the ΔLOD, GPS and geomagnetic data have a high degree of consistent synchronicity, and none of them has a stable decreasing trend. Based on the AR-z spectrum method (Ding and Chao, 2015b, 2018b) and upon using a much longer ΔLOD time series (1760-2018), D19 identified 9 periodic signals, i.e., the ~149yr, ~68yr, ~33yr, ~22.3yr, ~18.6yr, ~13.5yr, ~11yr (or said the ~10.6yr signal; which has also been found in the $J_2$ time series; see Ding and Chao (2018b) and Chao et al. (2020)), ~8.5 yr and ~5.85 yr periodic signals, in the intradecadal and decadal ranges. They first found that the ~8.5yr periodic oscillation (EYO) and suggested that it can be represented by a stable *cosine* signal. In D19, a clean time series for the SYO was obtained (after fitting and removing the other periodic signals; a similar process to that in Holme and de Viron (2013)). D19 showed that the SYO has no stable decay trend in the 1962-2018 time span, a slight decreasing trend in the 1975-1995 time span, but an increasing trend after 1995; this finding is consistent with the result shown of Chao et al. (2014), and D19 explained this by a modulation. Based on Daubechies wavelet fitting, NMWT, and a BEPME (boundary extreme point mirror-image-symmetric extension) method (which was claimed to avoid the edge effects in the NMWT), Duan and Huang (2020a) (referred to as DH20) analyzed the ΔLOD in the 1962-2019 time span. Their results confirmed that the SYO has a stable decreasing trend and that the EYO has a stable increasing trend, and they still suggested that the SYO has a $Q$ of ~51.

Regarding the possible relationship between the SYO/EYO and the geomagnetic jerks, Holme and de Viron (2005) first found that the 1969, 1972, 1978, 1982, 1992 and 1999 jerks are consistent with the sudden changes in the ΔLOD in the 1962-2005 time-span. Holme and de Viron (2013) further confirmed that the sudden changes (jumps) in their cleaner SYO time series may be triggered by geomagnetic jerks. Silva et al. (2012) also suggested that the SYO seems to be closely related to some geomagnetic jerks. Chulliat et al. (2015) suggested that the SYO may be related to the 2006, 2009 and 2012 jerks. Soloviev et al. (2017) also suggested that the SYO has some relationship with the 1996, 1999, 2002 and 2014 geomagnetic jerks. However, by using the modelled torsional waves, Cox et al. (2016) found that ~6yr period torsional oscillation cannot produce observed jerk signals. D19 first calculated the excitation function time series $\varphi(t)$ for the SYO in ΔLOD based on a deconvolution process. They found that there is no clear relation between the geomagnetic jerks and $\varphi(t)$, but they concluded that this finding needs to be confirmed by further work. Duan and Huang (2020b) claimed that they did not identify any possible excited event for the SYO since 1962 and suggested that the SYO is discontinuously excited with a random 50-100yr time interval. Upon using 13 selected jerks, DH20 found that the peaks/valleys of the EYO are consistent with 10 of them, but they also suggested that the SYO has no such relationship with jerks. DH20 further concluded that the EYO can be used to predict jerks.

To date, there are two disputes for the SYO and the EYO in the ΔLOD:

1) Whether the SYO and the EYO have had stable damping trends since 1962.

2) Whether the geomagnetic jerks are possible excitation sources for the SYO and the EYO.

In this study, we try to resolve those two disputes.

## 2. Whether the SYO and the EYO have stable damping trends

In this section, we will use two ΔLOD time series with different lengths as the datasets, and the ΔLOD time series in Holme and de Viron (2013) and DH20 will be extracted for further use.

### 2.1 The used datasets

We first extract the residual ΔLOD time series ($R_0$; 1962/01-2018/05; daily sampling) and the recovered SYO/EYO ($S_1/E_1$) from DH20 (see Supplementary Figure S1 for the comparisons between the original figures of DH20 and our extracted results); the SYO time series $S_2$ (1962/01-2011/06; monthly sampling) from Fig. 2 of Holme and de Viron (2013) is also extracted for further comparison.

Furthermore, we choose the 1962-2020 ΔLOD time series from the EOPC04 dataset (Petit & Luzum, 2010), the atmospheric angular momentum (AAM) dataset, the oceanic angular momentum (OAM) dataset and the hydrological angular momentum (HAM) dataset for comparison and further use (the AAM/OAM/HAM are the main Earth external excitation sources of the Earth's rotation). The 1962-2020 ΔLOD record is first decimated from 1 day to 10 days sampling (we first use a low-pass filtering with an 18.263cpy cut-off frequency to filter the original time series, and then resample it from 1 day to 10 days; the similar manner is used in the following). The AAM record (1948/01-2019/03) (mass terms + motion terms) is also decimated from 6 hours to 10 days sampling. The OAM records (mass terms + motion terms) are combined by two different datasets, ECCO_50yr.oam and ECCO_kf080i.oam; the timespan of the first dataset is 1949/01/06-2003/01/06 and the sampling is 10 days; the timespan of the second dataset is 1993/01/02-2019/02/15, and the sampling is daily. We merged them to form a new record with a 1949/01-2019/02 timespan and a 10-day sampling interval. The HAM record is downloaded from the Special Bureau for Hydrology (based on the Land Surface Discharge Model (LSDM)) (Dill, 2008), the timespan is 1971/01/01-2020/06/04 and the sampling is daily; again, it is decimated from daily to 10 days sampling. The ΔLOD and the AAM/OAM/HAM excited LOD time series are shown in Fig. 1a (here, we note that the units in the figures of D19 were misspelled to *mas*), and their corresponding Fourier amplitude spectra are shown in Fig. 1b. We can see that the ΔLOD shorter than 5yr are mainly caused by the AAM effects. All the AAM, HAM and OAM effects have very small contributions to the intradecadal period band (5-10yr; i.e., the 0.1-0.2 cpy frequency band), except that the ~5yr signal in the ΔLOD is caused by the AAM (see Figure 1b). Hence, here, we only remove the AAM effect from the EOPC04 ΔLOD time series. The residual ΔLOD time series after removing the AAM effect and the residual seasonal and tidal effects is referred to as $G_1$ (=ΔLOD–AAM).

As the EOPC04 ΔLOD time series only has an ~58yr length, which may be too short to isolate two close signals by using a filter, we further choose a yearly long-term ΔLOD time series (1730-2020) (from: www.iers.org/IERS/EN/Science/EarthRotation/LODsince1623.html?nn=12932).

Before we do further works, we list the time series used in the following sections and their corresponding sources/preprocessing process in Table 1.

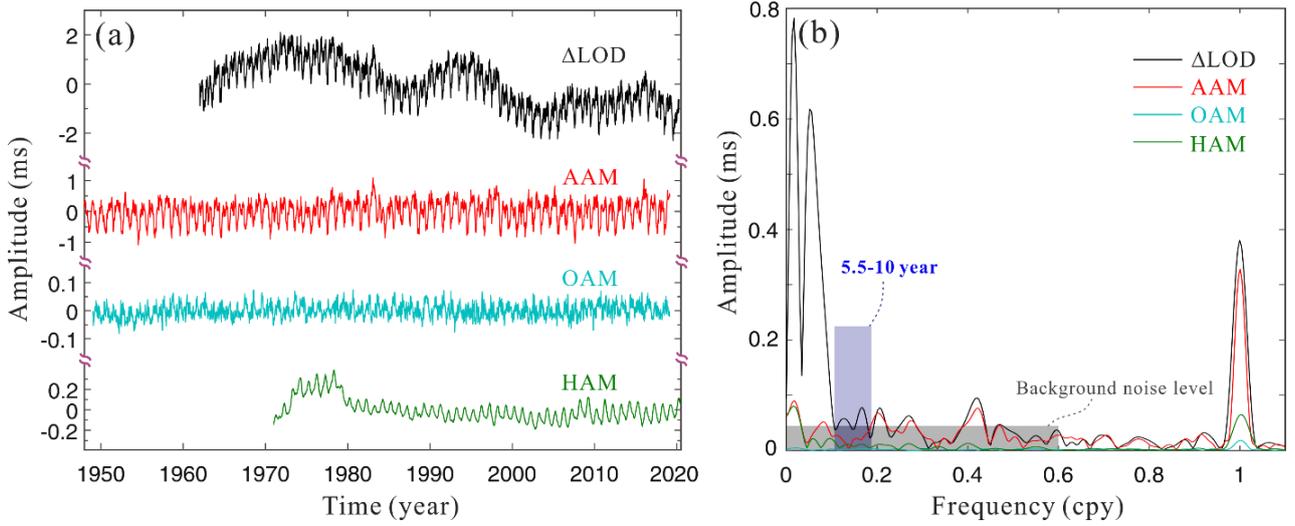

**Figure 1**. (a) The ΔLOD and the AAM/OAM/HAM excited LOD time series. (b) The corresponding Fourier spectra of the time series in (a). The gray area denotes the background noise level in the 0-0.6cpy, and the blue area denotes the 5.5-10yr period band (0.1-0.182 cpy frequency band).

**Table 1**. The used time series in this study and their corresponding sources/preprocessing process

| Name | Sources/Preprocessing process |
|---|---|
| $R_0$ | The extracted residual ΔLOD time series from Fig. 1a of DH20. |
| $R_1$ | The filtered $R_0$ after applying a zero-phase high-pass filter (with a 0.14 cpy cut-off frequency). |
| $S_1$ | The extracted SYO time series from Fig. 2a of DH20. |
| $E_1$ | The extracted EYO time series from Fig. 2b of DH20. |
| $S_e$ | An ~7yr oscillation obtained by using the NMWT+BEPME to the residual time series $R_0-(S_1+E_1)$. |
| $G_1$ | The residual ΔLOD time series obtained from removing the AAM effects from the EOPC04 ΔLOD time series (the residual seasonal and tidal signals also have been fitted and removed). |
| $S_0$ | A high-pass filter (with 0.145 cpy as the cutoff frequency) is firstly used to $G_1$, and then a 6-month running mean is used. |
| $S_2$ | The extracted SYO time series from Fig. 2 of Holme and de Viron (2013). |
| $E_0$ | Obtained from the 1730-2020 ΔLOD time series after using a bandpass filter with 0.109cpy and 0.124cpy as the cut-off frequencies. |
| $G_2$ | Obtained from $G_1$ after using a high-pass filter with 0.10cpy as the cut-off frequency. |
| $S_3$ | The SYO time series obtained from $G_2$ after applying a bandpass filter (the cutoff frequencies are 0.145 cpy and 0.205 cpy). |
| $Se_1$ | The fitted 7.6yr signal from $G_2$ |

## 2.2 Reanalysis of the results in previous studies

The extracted $R_0$, $S_1$ and $E_1$ and their corresponding Fourier spectra are shown in Figs. 2a-2d (also see Supplementary Fig. S1). The spectra of $S_1$ and $E_1$ show clear differences from that of $R_0$, and the spectrum of the residual $R_0$–($S_1$+$E_1$) clearly shows a residual peak between the two target signals (see Fig. 2d; located at ~0.143cpy, i.e., ~7yr period). The phases also show clear differences in the target frequency bands (Figs. 2e and 2f). These results preliminarily indicate that the recovered SYO/EYO in DH20 cannot completely represent the real signals in ΔLOD.

When using the classical filtering method for isolating two close signals, the data length should preferably be longer than $2/\Delta f$ (where $\Delta f$ is the frequency interval). As $R_0$ is ~56yr long, one can safely use a classical high-pass filter to isolate the SYO, but it is not possible to completely filter the ~10.6yr signal (Fig. 2d) for the EYO from $R_0$. Here, a residual $R_1$ is obtained after applying a zero-phase high-pass filter (with a 0.14 cpy cut-off frequency) to $R_0$ (see Fig. 2b, the black curves). The spectral results show that the SYO amplitudes in $R_0$ and $R_1$ are almost the same (considering the filtered noise effects; see Fig. 2d) and that the phase difference between them is approximately zero (Figs. 2e and 2f). These findings indicate that the applied filter does not change the real SYO in $R_0$. After using a high-pass filter (with 0.145 cpy as the cutoff frequency) for $G_1$, we further apply a 6-month running mean as done by Holme and de Viron (2013), then we can obtain a residual time series which almost only contain the SYO signal, we refer it as to $S_0$. From Fig. 2b, we can find that

$S_0$, $S_2$ and $R_1$ have good consistency even though different methods were used (D19; Holme and de Viron, 2013), and there is no stable decreasing trend for the SYO in the 1962-2019 timespan. In fact, $S_1$ from DH20 shows consistency with $R_1$ and $S_0$ only in the 1970-2000 timespan (simulation tests also prove this; see Supplementary Fig. S2).

For the EYO, a bandpass filter was applied to the 1730-2020 ΔLOD time series. In the 1962-2019 time-span (see Fig. 2c), the EYO is almost a stable oscillation, which is consistent with the finding in D19.

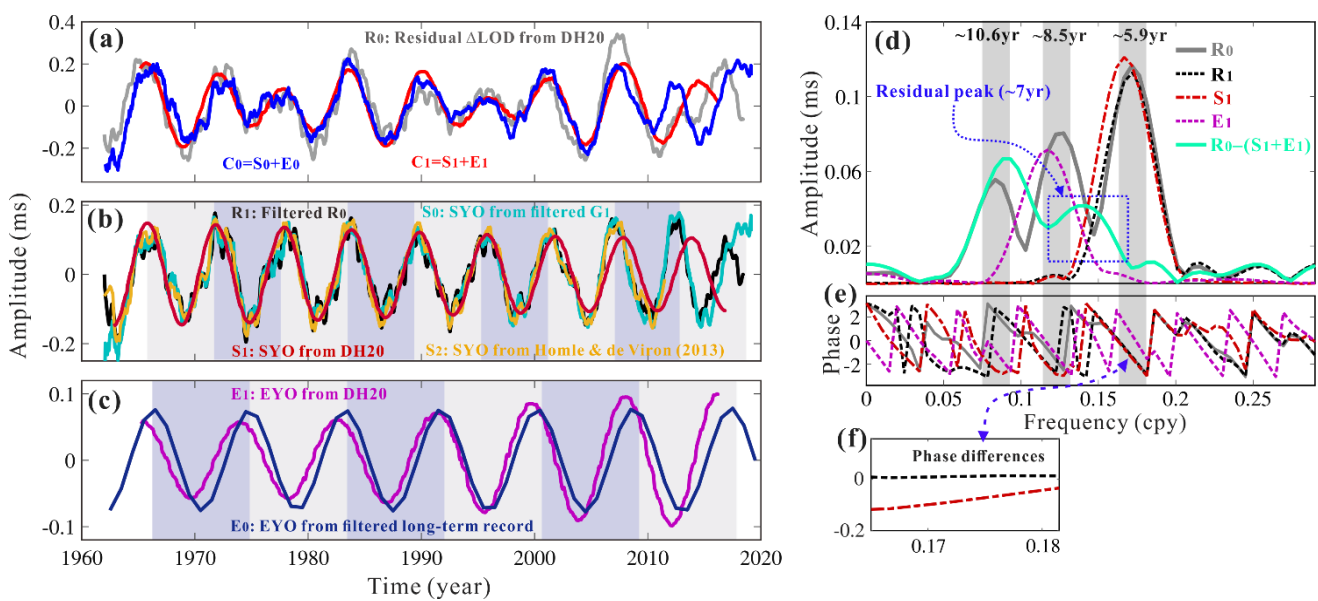

**Figure 2.** (a) The residual ΔLOD time series ($R_0$, gray curve) extracted from DH20; (b) $R_1$ (a filtered version of $R_0$), the extracted SYO ($S_1$) time series extracted from DH20, $S_0$ (SYO from the filtered $G_1$) and $S_2$ (SYO extracted from Holme and de Viron (2013)); (c) the extracted EYO ($E_1$) time series extracted from DH20, and $E_0$ (EYO obtained from a long-term ΔLOD record); $C_0=S_0+E_0$ and $C_1=S_1+E_1$ are also plotted in (a). (d) and (e) show the corresponding Fourier amplitude and phase

spectra of $R_0$, $R_1$, $S_1$, $E_1$ and $R_0-(S_1+E_1)$. (f) The phase differences between $R_1/S_1$ and $R_0$ in a narrow frequency band.

The above reanalysis denotes that the SYO and EYO obtained by DH20 cannot completely represent the real signals in ΔLOD, and there seems to also be an ~7yr signal in the 5.5-10yr period band of $R_0$. To confirm this, we further used the same process (NMWT+BEPME) to further obtain an ~7yr signal from the residual time series $R_0-(S_1+E_1)$, and a stable increasing time series $Se(t)$ was obtained (see Fig. 3a). This is a strange finding, all the three close signals have stable damping trends, and it is difficult to image what physical mechanisms can cause such observations. Fig. 3b shows that the restored ~7yr signal can well represent the residual spectral peak. However, the three damping signals ($S_1+E_1+Se$) cannot well represent the original $R_0$ in the corresponding frequency band, neither for the amplitudes nor for the phases (see Fig. 3b). To further confirm this, we also show the NMWT spectrum of $R_0-(S_1+E_1)$ in Fig. 4, which clear shows that there are still some residual energy for the ~5.9yr and ~8.5yr periods, and there is no stable signal for the ~7yr period. Hence, we can sure that the residual peak around ~7yr period in Fig. 2d is mainly caused by the residual energy from the ~5.9yr and ~8.5yr signals (because those two obtained damping oscillations cannot well represent the original signals in the residual ΔLOD).

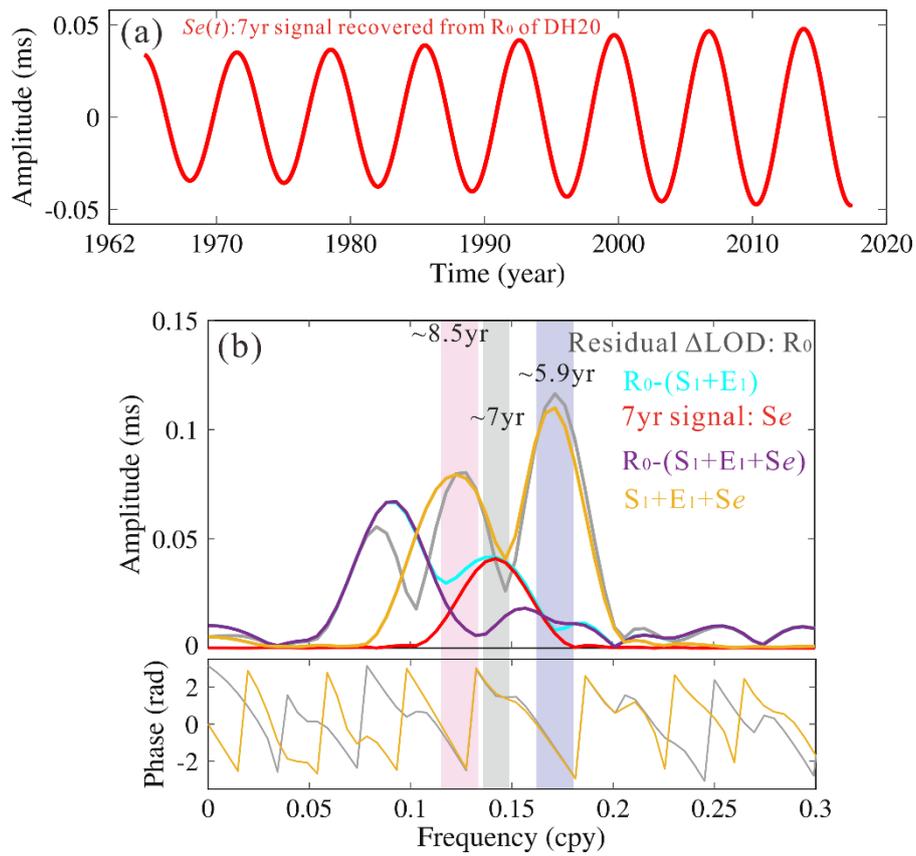

**Figure 3.** (a) The recovered 7yr signal from the exacted residual time series $R_0$ after removing the SYO ($S_1$) and EYO ($E_1$) based on the same process used in DH20. (b) The amplitude and phase spectra of $R_0$, the restored 7yr signal $Se$, $R_0$–($S_1$+$E_1$), $R_0$–($S_1$+$E_1$+$Se$) and $S_1$+$E_1$+$Se$.

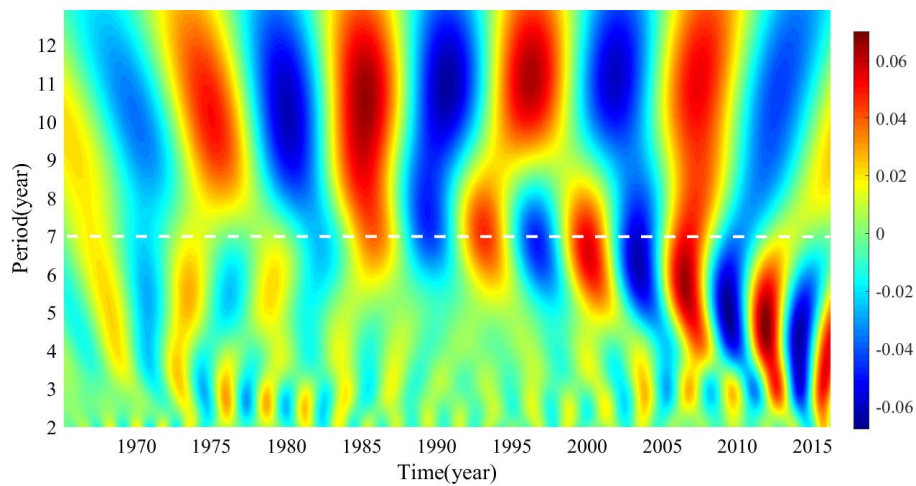

**Figure 4.** The NMWT spectrum of the residual time series $R_0-(S_1+E_1)$ which is obtained after removing $S_1$ and $E_1$ from $R_0$.

**2.3 New results from the classic filter**

We first apply a high-pass filter to the residual ΔLOD time series $G_1$ ($G_1$=ΔLOD−AAM; the cutoff frequency is 0.10cpy), and refer it as $G_2$; the Fourier spectrum of $G_2$ is shown in Fig. 5b (and 5c; black curve). Compared with Fig. 2d, we can find that the ~10.6yr signal has been filtered. This ~10.6yr signal may original from the solar cycle, but further confirmations still need (see Chao et al. 2020). Here we note that, in D19, a 7.7yr signal has been fitted and removed to obtain a cleaner SYO time series; given that the amplitude of this signal is quite small, this information was not specifically explained in D19. In the following, we will also show that there is actually an ~7.6yr signal in the ΔLOD, not an ~7yr signal. As the length of $G_2$ is approximately 58yr, even though there is an ~7.6yr signal, it is still long enough to filter the SYO from the EYO and the ~7.6yr signal, but it is difficult to isolate the EYO with an ~7.6yr signal. Hence, we use a bandpass filter (the cutoff frequencies are 0.145 cpy and 0.205 cpy) to isolate the SYO from $G_2$ (see Fig. 5a), we refer this SYO time series as $S_3$; while the EYO is selected from the filtered result ($E_0$) that from the 1730-2020 ΔLOD time series (i.e., the time series $E_0$ in Fig. 2c; also plotted in Fig. 5a). We did not try to filter the ~7.6yr signal from the 1730-2020ΔLOD time series, because the amplitude of ~7.6yr seems too small (only ~0.025ms) and the data before 1962 are too noisy. Instead, we first remove the obtained $S_3$ and $E_0$ from $G_2$ and then use iterative fitting to obtain the ~7.6yr signal. Finally, we find that a 7.6yr periodic signal can well represent this signal (refer it as *Se*$_1$, see Fig. 5a; green curve). From Fig.

5a, we can clearly find that the SYO and EYO have no stable damping trends. The EYO is close to a stable oscillation, while the SYO only has a decay trend in the 1962-1990 time span but has an increasing trend in the 1990-2008 time span; the whole envelop of the SYO seems a modulation phenomenon (see the envelopes denoted by the dashed curves in Fig. 5a). Fig. 5b shows the Fourier amplitude and phase spectra of $G_2$, SYO+EYO and the residual time series ($G_2$–(SYO+EYO)). The amplitudes and phases around the ~5.9yr show that the obtained SYO can well represent the original spectra, but the results around the ~8.5yr show a clear difference between the EYO and the original signal. Similar to Fig. 2d, there is a residual peak around the ~7.6yr. Fig. 5c is similar to Fig. 5b, but the fitted 7.6yr signal $Se_1$ is added into $S_3+E_0$ time series. The results show that the three obtained time series can well represent the original signals in the 5.5-10yr period band (i.e., the 0.1-0.182cpy frequency band; not only for the amplitude, but also for the phase). However, we must say that the amplitude of this ~7.6yr signal is too small to claim that it must be a 'signal'. To further confirm whether this ~7.6yr signal is a periodic signal or not, we try to search for this signal in the global geomagnetic observations. The optimal sequence estimation (OSE) method (see Ding and Shen (2013) and Ding and Chao (2015a) for the details) is used for such purpose.

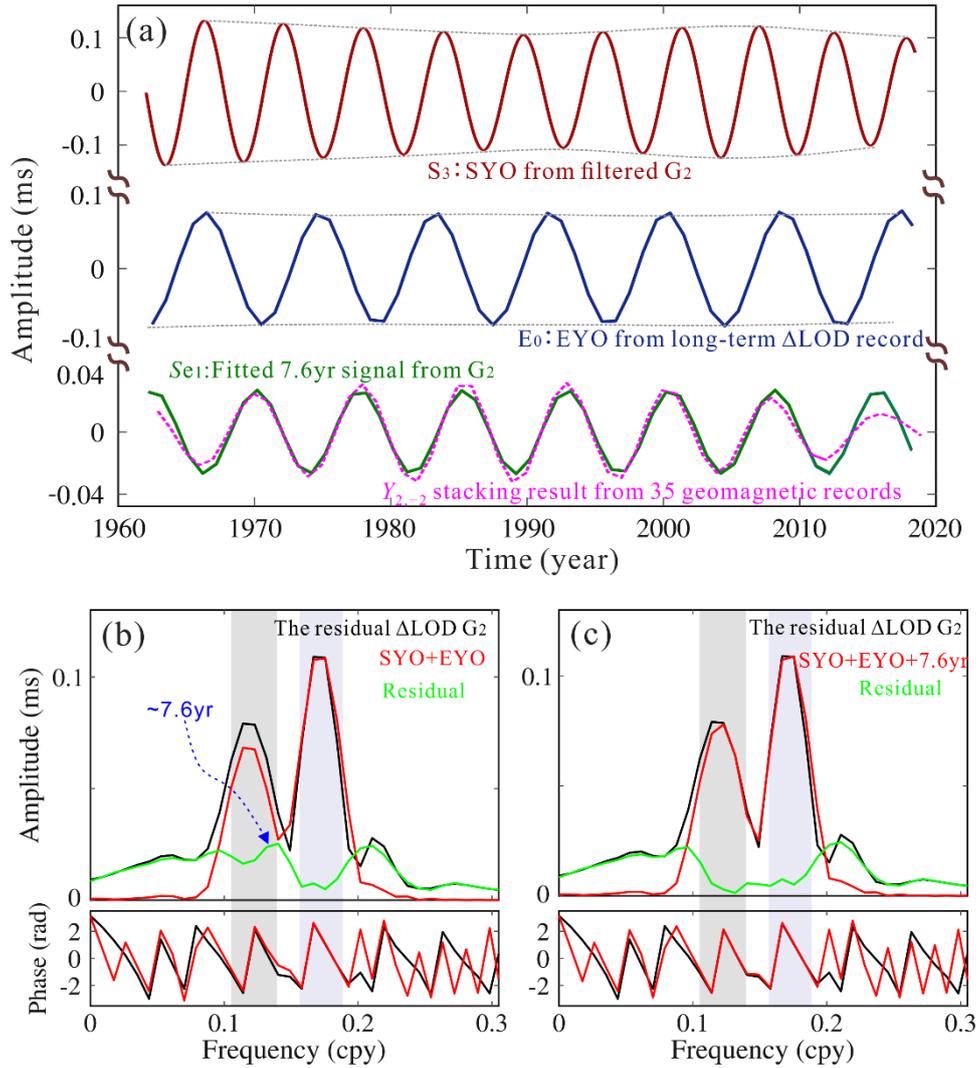

**Figure 5**. (a) The obtained SYO ($S_3$) and EYO ($E_0$) from the ΔLOD time series based on the classic filter process; the green curve denotes the fitted 7.6yr signal from the $G_2$ time series (the filtered $G_1$ with a 0.1cpy cut-off frequency) after removing the SYO and EYO in (a). (b) shows the amplitude and phase spectra for $G_2$ and the SYO+EYO (in (a)); the amplitude spectrum for the residual time series ($G_2$–(SYO+EYO)) is also shown in (b). (c) is similar to (b), but further considers the fitted 7.6yr signal.

Inasmuch as the OSE method needs the used records have same time spans, we finally selected radial components ($B_r$) of 35 surface geomagnetic records (we have converted the provided geomagnetic intensity $F$ and inclination $I$ into the radial component by $B_r = -F \sin I$) with have common 1962/01-2019/05 time span. We first use a band-pass filter (with 0.105cpy and 0.15 cpy as the cut-off frequencies) to those time series, and then use OSE to those filtered records. Finally, we find that there is a ~7.6yr periodic signal which is only presented in the obtained $Y_{2,-2}$ spherical harmonic related time series. The whole process is similar as the process used for the SYO in Ding and Chao (2018a), here we will not show the details. The normalized $Y_{2,-2}$ related stacking time series is showing in Fig. 5a (the dashed purple curve), we can find that the ~7.6yr signal in ΔLOD and geomagnetic data has a high degree of consistent synchronicity. Hence, we may conclude that the obtained SYO, EYO and the 7.6yr signal are reasonable.

Here we want to further note that the two statements, '*whether the SYO and EYO have decreasing or increasing trends in the time domain observation*' and '*whether the SYO and EYO are attenuation oscillations*', are different. The former is focus on the fluctuation characteristic of the SYO and EYO in the time domain, while the latter is focus on the quality factor $Q$ of the SYO and EYO. Generally, if the $Q$ of an oscillation is larger than zero, this oscillation is a decay oscillation, otherwise the oscillation is an increasing oscillation. In the Earth system, almost all the well-known oscillations or normal modes have positive $Q$ values. In the time domain (or said in the real observations), decreasing or increasing characteristic is not necessary for an attenuated oscillation in the time domain, for example, the Chandler wobble has a ~20-170 $Q$ value (see as Ding and Chao (2017)), but its amplitude is time-vary rather than damped in the time domain. If an oscillation is continuously excited, it seems no available method can directly estimate its $Q$ value in the time domain (similar as the case for the Chandler wobble). The above analysis are only focus on that the

SYO and EYO have no damping trend in the time domain, we will try to estimate the $Q$ values for the SYO and EYO in Section 4.

**3 Why were the decay trends for the SYO and EYO obtained?**

The above sections show that all the results from the classic filter and from Holme & de Viron (2013) have no stable trend for the SYO and EYO. Therefore, why do the SYO and EYO recovered in DH20 have stable damping trends? Actually, Supplementary Figs. 3, 9-10, 14 and 15 in DH20 clearly show that the edge effects were still present even when the NMWT+BEPME method was used. More importantly, their Figs. S14a and S15a show that the recovered signals have clearly increased and slightly decreased amplitudes, respectively, even the input signals are stable *sine* signals. In light of this, we suspect that the damping trends for their SYO and EYO results should be affected by the methods that they used.

We reproduced the same processing strategy explained in DH20 (Daubechies wavelet fitting+NMWT+BEPME) and tested it. Two time series, $S_1(t)$ and $S_2(t)$, were simulated. $S_1(t)$ contains 9 zero-phase and stable *cosine* signals (the same as those in DH20, except for the random noise term; see their Supplementary Information); $S_2(t)$ also contains the same 9 periodic signals, but the amplitudes and phases were estimated by using a least-square process to fit the observed ΔLOD (see Fig. 6a; dark blue curve). The SYO/EYO restored from $S_1(t)$ (+random noise) are shown in Fig. 6b (green curves), and our restored results are almost the same as those in DH20 (red curves in Fig. 6b) (considering that the random noise cannot be the same). We further used the same process to reanalyze $R_0$, and the restored SYO/EYO curves (green curves in Fig. 6c) almost coincide with the

SYO/EYO curves extracted from DH20 (red curves in Fig. 6c), slight differences arise from the errors introduced when extracting data from their figures. Figs. 6b and 6c prove that we have fully reproduced the processing strategy used in DH20.

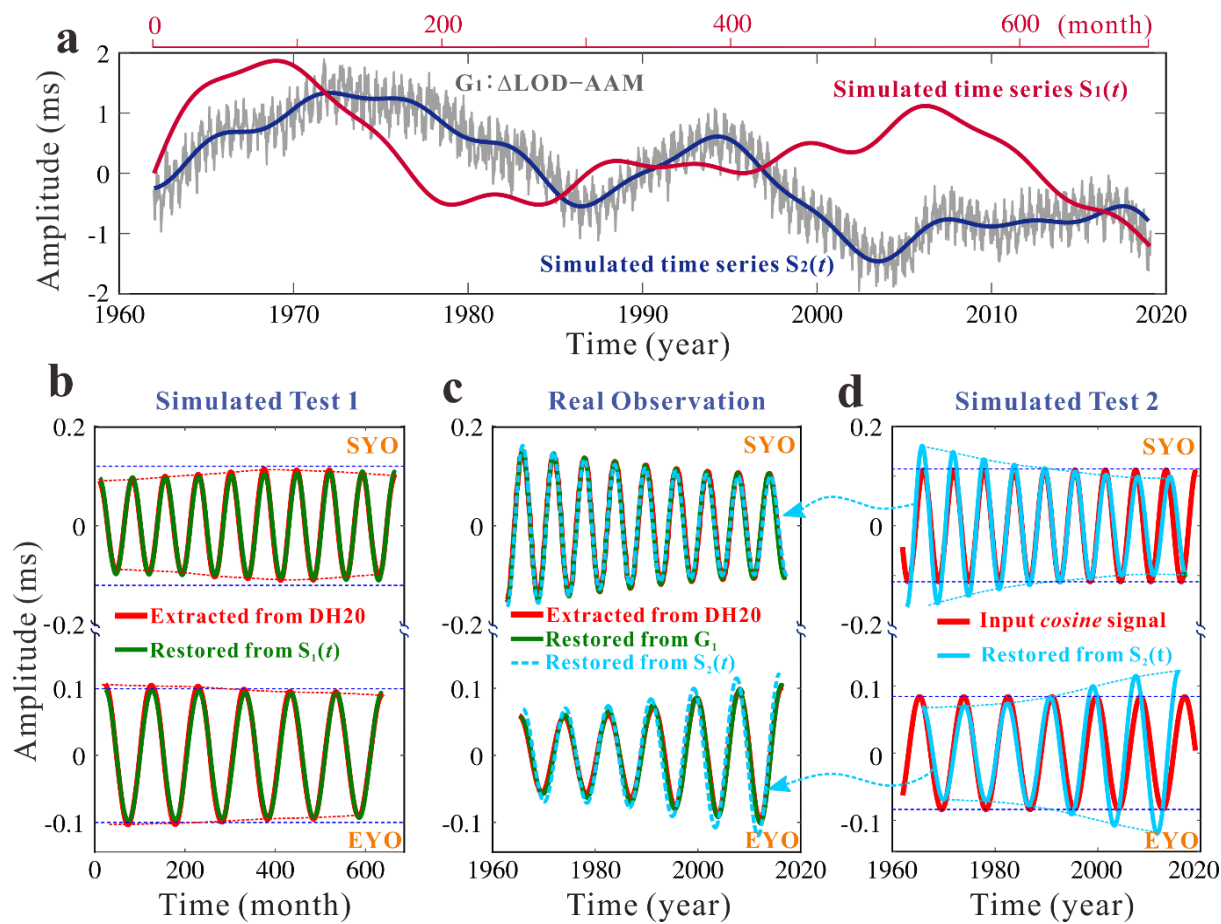

**Figure 6.** (a) The residual time series $G_1$ (ΔLOD–AAM) and two simulated records $S_1(t)$ and $S_2(t)$. (b) The restored SYO and EYO from $S_1(t)$ (green curves) based on the same processing strategy used in DH20 (Daubechies wavelet fitting+NMWT+BEPME); the extracted SYO and EYO from Figs. S14a/S15b in DH20 (red curves) are also shown in (b). (c) The restored SYO and EYO from $G_1$ (green curves) based on the same process used in (b); and the extracted SYO and EYO from DH20 (also shown in Figs. 2b and 2c). (d) The input SYO/EYO (red curves) and the restored SYO/EYO from $S_2(t)$. The restored SYO/EYO from $S_2(t)$ are also plotted in (c) for comparison. The blue dashed

lines in (b) and (d) denote the input fixed amplitudes, and the colored dashed curves denote the corresponding envelops.

Given these results, we used the same process to analyze $S_2(t)$. Not surprisingly, we obtained a decreasing SYO and an increasing EYO for the input stable *cosine* signals (see Fig. 6d). As the real observation is more complicated than $S_2(t)$, there is a possibility that even if the SYO/EYO are nearly stable, the NMWT+ BEPME will indicate SYO/EYO damping trends. We re-plot the SYO/EYO restored from $S_2(t)$ in Fig. 6b to compare with the SYO/EYO restored from the real observed ΔLOD $G_1$, the two results are well consistent; which means that our simulated *cosine* SYO and EYO may very close to the real signals in the ΔLOD. Here, we may conclude that the damped nature of the SYO/EYO was only an artifact of the method used in DH20 (more evidence can be obtained from the test codes in the Supplementary Information).

## 4. The relationship between jerks and EYO/SYO

### 4.1 Comparing jerks with the SYO/EYO in the time domain

As we have reviewed in section 1, whether the SYO/EYO have certain relationships with jerks is debatable, although we tend to believe they do have some relationships. In this section, we first follow the thoughts outlined in DH20, i.e., we compare the peaks/valleys of the SYO and the EYO with the jerks, but we will consider more jerks.

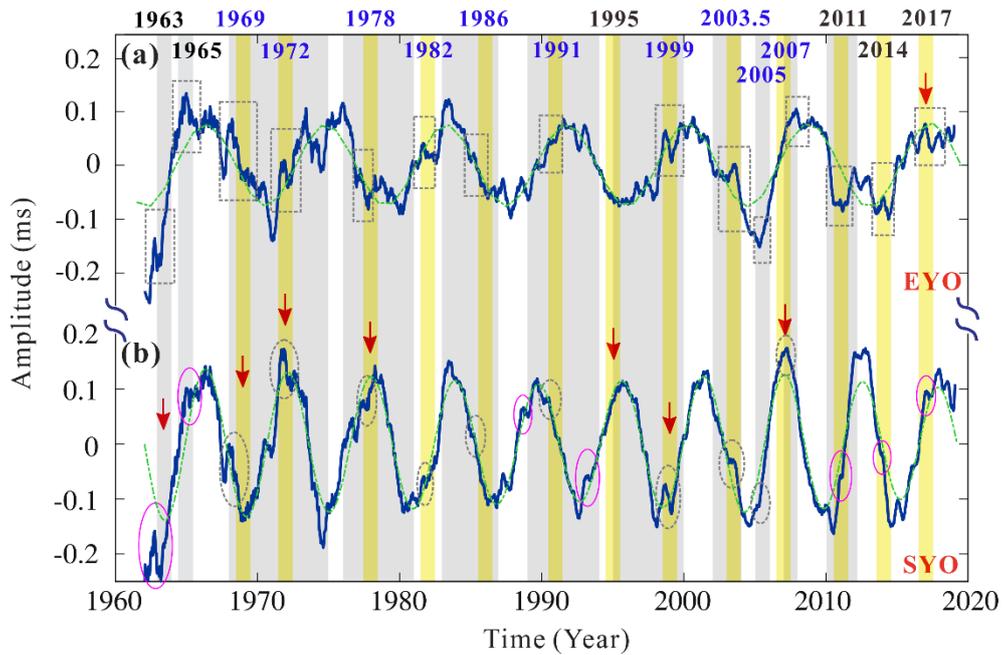

**Figure 7**. The EYO (a, dark blue curve) and SYO (b, dark blue curve; same as $S_0$ in Fig. 2b) obtained from $G_2$ after a six month smooth process and the jerk events in recent decades; the green curves denote the obtained EYO and SYO which have shown in Fig. 5a. The yellow areas denote the 13 jerks (with one year uncertainty) selected in DH20, some of them were also used in Holme and de Viron (2013) (the corresponding year is marked with blue font at the top of (a)). The gray areas show the jerk bounds from 1957 to 2008, and a new 1965 jerk is added. In (a), the dashed gray rectangles indicate the sudden changes in the EYO which may correspond to the jerks. In (b), the dashed gray ellipses indicate the sudden changes in the SYO which may correspond to the jerks used in Holme and de Viron (2013); and the purple ellipses indicate the sudden changes in the SYO which may correspond to the other jerks.

In DH20, only 13 jerks were selected (see the yellow areas in Fig. 7), although many jerks have been identified in the past decades. Even don't consider the jerk bounds reviewed in Brown et al. (2013) in the 1968-2010 time span, at least the 1963 and 1965 jerks were missed in their study (see the gray areas in Fig. 7). For a given jerk, it will not occur at the same time in different regions of the Earth surface, and it has 1-2yr uncertainty (Brown et al., 2013; Pinheiro et al., 2013; Chulliat & Maus, 2014). In Fig.7, we show two residual time series for the EYO and SYO, respectively. For the EYO time series, we first remove the obtained SYO ($S_3$; see Fig. 5a) and the 7.6yr signal ($Se_1$; see Fig. 5a) from $G_2$ (the filtered $G_1$ with a 0.1 cut-off frequency). After then, we use a six month smooth to the residual time series (similar as we have done for the $S_0$ time series in Fig. 2b); the final smoothed residual time series is shown in Fig. 7a. For the SYO time series, we directly use the $S_0$ time series which has been shown in Fig. 2b (also see Fig. 7b). Meanwhile, in Fig. 7a and 7b, we also respectively plot the EYO ($E_0$) and SYO ($S_3$) time series which have shown in Fig. 5a (see the dashed green curves in Fig. 7).

The 13 jerks used in DH20 were indicated by the yellow areas in Fig. 7, some of them were also used in Holme and de Viron (2013) (see the corresponding years which were marked with blue font at the top of Fig. 7a). Meanwhile, the gray areas in Fig. 7 denote the geomagnetic jerk bounds (and jerks) which referred from Brown et al., (2013) (before 2010), Chulliat et al. (2015) and Torta et al. (2015) (after 2010). Comparing the peaks/valleys of the SYO and EYO with the 13 jerks and the 1963 and 1965 jerks, different with the findings in DH20, we find that only the 2017 jerk is well consistent with the peak of the EYO (see the red arrow in Fig. 7a); while there are seven jerks are well consistent with the peaks/valleys of the SYO (see the red arrows in Fig. 7b) (including the best known, and most studied 1969 and 1978 jerks; see also in Fig. 3 of Holme and de Viron (2013)), but DH20 suggested that there is no such consistency between the jerks and the peaks/valleys of the SYO. Until now, there has no robust evidence to prove that the EYO must be caused by a torsional oscillation in the fluid outer core, although DH20 has suggested this (as Duan et al. (2018) have suggested for the SYO). If the EYO is increased by jerk excitations, as claimed in DH20, the SYO should also be increased by jerk excitations (contradict with the decreased result obtained by DH20). Considering that different jerks generally have 1-3yr intervals and there is only the 2017 jerk that corresponds to the peak of the EYO, the EYO seems to offer no remarkable help in predicting jerks.

According to Cox et al. (2016) and Pinheiro et al. (2019), jerks may be somewhat smoother than what is often considered, and Pinheiro et al. (2019) found that the jerk duration time can be up to 4.7yr; hence, even have done the above, we do not recommend such comparison. More reasonable comparisons should be determined by comparing the sudden changes in the SYO/EYO time series with jerks (similar that done by Holme and de Viron (2013)) or comparing their excitation sequences with jerks (similar that done by D19).

In Fig. 7b, the dashed gray ellipses indicate the sudden changes in the SYO which may correspond to the jerks used in Holme and de Viron (2013); we can see that only the 1995 jerk has no clear sudden change in the SYO time series. The purple ellipses in Fig. 7b indicate the sudden changes in the SYO which may correspond to the other jerks. In Fig. 7a, it seems that all the 15 jerks have corresponding sudden changes in the EYO (the dashed gray rectangles). Based on the results in Fig. 7, we may conclude that the jerks seems do correspond to the sudden changes in the SYO and EYO time series.

However, we also note that it is not easy to determine 'sudden changes' in the SYO and EYO time series, there are also probably just caused by high frequency noise. To further confirm the possible relations, we will compare the excitation time series of the SYO and EYO with jerks.

**4.2 Comparing jerks with the excitation time series of the SYO/EYO**

Before we calculate the excitation time series of the SYO and EYO, we first try to estimate the *Q* values of them.

As we have confirmed that the SYO and EYO don't caused by the Earth external excitation sources (see Fig. 1), we can reasonable assume that they must be caused by some core motions (of course the more detailed mechanism need to be further studied in the future). Namely, the SYO and EYO in the ΔLOD can physically be treated as normal modes of the Earth, hence they can be equal to the temporal convolution of harmonic oscillations with the 'excitation function' $\varphi(t)$ (*Chao*, 2017b):

$$m(t) = -i\omega e^{i\omega t} \int_0^t e^{i\omega\tau} \varphi(\tau) d\tau. \tag{1}$$

where $\omega=2\pi/P(1+i/2Q)$, *P* is the period. Here a deconvolution method can be used to estimate *P* and *Q* values of the SYO and EYO. This deconvolution method has been successful used in Furuya and Chao (1996), Gross (2005), Chao and Chung (2012) for the Chandler wobble, in Chao and Hsieh (2015) for the free core nutation, and in D19 for the SYO. The detailed introduction about this method can be found in Chao and Chung (2012) and D19, here we only given a short introduction.

Suppose the true values of *P* and *Q* are not known in advance, we can choose a set of [*P*, *Q*] as the input parameters for the deconvolution. If the used [*P*, *Q*] has deviations with the true values, then the deconvolution notch filter would then be somewhat off the target and unable to remove the resonance power completely (Chao and Heish, 2015). Therefore, some extra power will be still present in the obtained excitation function $\varphi(t)$. If the excitation is statistically independent of the observation noises, the spectral power around the target frequency bin will be significantly improved. Given this, if a deconvoluted $\varphi(t)$ has the least power in the target frequency band, the corresponding input [$P$, $Q$] will correspond to the correct values.

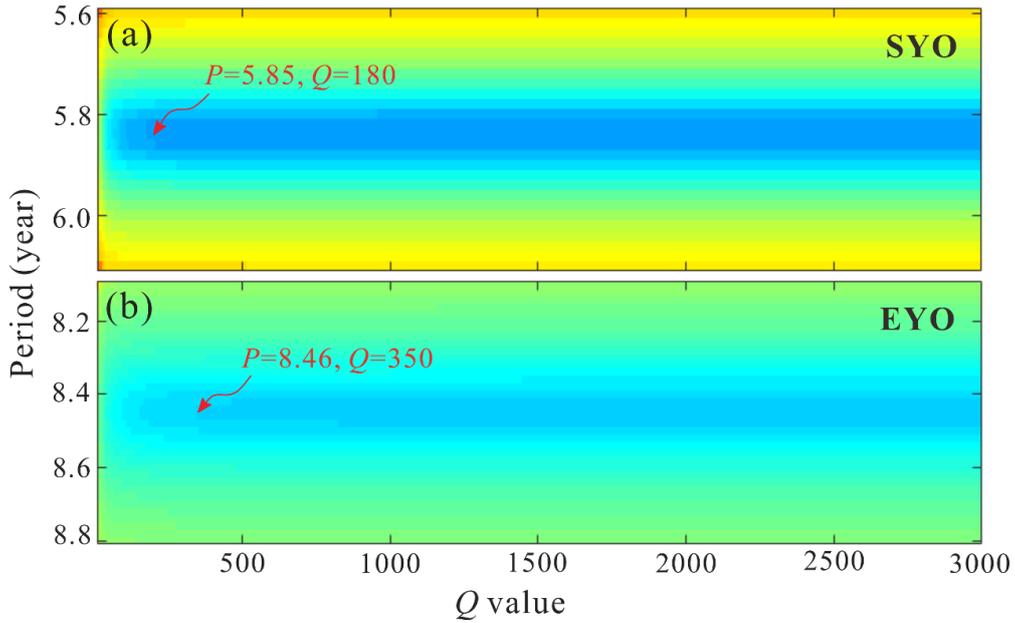

**Figure 8**. The mean excitation power of the excitation $\varphi_S(t)$ (a: for SYO) and $\varphi_E(t)$ (b: for EYO) as a function of the input [$P$, $Q$].

Here we set $P$ from 5.6yr to 6.1yr and from 8.1yr to 8.8yr for the SYO and EYO, receptively; the period interval is 0.02year for both of them. And we set $Q$ from 1 to 3000 with 10 interval for both of them. As for the chosen SYO and EYO time series, to more generous, we choose the same time series used in Fig.7 but without a six month smooth. After the deconvolution process, the resultant mean excitation power as a function of the input [$P$, $Q$] are shown in Fig. 8. The results show that no minimum can be determined in Figs. 8a and 8b, but around $P \approx 5.85$ and $Q \geq 180$ in Fig. 8a and around $P \approx 8.46$ and $Q \geq 350$ in Fig. 8b, the mean excitation powers are clearly lower than the others, which mean that the $Q$s of SYO and EYO should be larger than 180 and 350, receptively. In D19, the $Q$ for the SYO was suggested should be larger than 200, considering that different lengths were used (1962/01-2016/07 in D19 and 1962/01-2019/02 in this study) and the estimate errors, such difference is acceptable. As for the more precisely periods, we finally obtain that $P=5.85\pm0.06$yr for the SYO (almost same as that given in D19: $5.85\pm0.03$yr) and $P=8.45\pm0.17$yr for the EYO.

After determining the $P$ and $Q$ for the SYO and EYO, we can further calculate their excitation function sequence according to Eq. (1). Here we use [$P=5.85$, $Q=180$] for the SYO and [$P=8.45$, $Q=350$] for the EYO; and the chosen SYO and EYO time series are same as those used for determining [$P$, $Q$]. The obtained excitation time series $\varphi_S(t)$ and $\varphi_E(t)$ for the SYO and EYO are shown in Fig. 9a, respectively. It can be found that $\varphi_S(t)$ and $\varphi_E(t)$ are almost overlapped with each other; given that the high-frequency (>2cpy) components of the chosen SYO and EYO time series are same as each other, this finding is easily understood. The $\varphi_S(t)$ (and $\varphi_E(t)$) has no very sharp waveform and no very significant spike, this means that the SYO and EYO should be continuously excited. Only from $\varphi_S(t)$ and $\varphi_E(t)$, it seems no more useful information can be obtained for the possible excitation sources; hence, we further use a 1D median filter (with a $n=18$ window) to the excitation time series $\varphi_E(t)$ to show some small perturbations (as $\varphi_S(t)$ and $\varphi_E(t)$ are almost same, so we can only do this for one of them). The obtained filtered $\varphi_{El}(t)$ is shown in Fig. 9b, different with the findings in D19, there are some sudden changes can be found in Fig. 9b (see changes around 1972, 1995 and 2014 for examples); this is mainly because a $n=105$ window for the 1D median filter was used in D19, this window is almost 3yr and too wide to show some local changes. To more clear show the sudden changes in $\varphi_{El}(t)$, we further calculate the first derivative $\varphi_{Ed}(t)$ of it, the result is shown in Fig. 9c. From the envelopes shown in Fig. 9c (yellow curves), the epochs of the first five maximum envelopes (except the envelope around 1967 which marked by the red arrow) are well consistent with the 1963, 1972, 1982, 1995 and 2005 jerks; and envelopes are generally well consistent with the jerks or jerks bounds (gray areas). Fig. 9d further shows the NMWT spectrum of the sequence $\varphi_{Ed}(t)$ in Fig. 9c. The spectral values in Fig. 9d are shown in the $\log_2$ scale that same as Alexandrescu et al. (1995, 1996) have done, if we follow the inferences of Alexandrescu et al. (1995, 1996), the jerks or jerks bounds (gray areas) in Fig. 9d seem to be well consistent with the sudden changes in $\varphi_{Ed}(t)$.

From the above analysis in sections 4.1 and 4.2, it seems that the geomagnetic jerks do relate to the SYO and EYO, although we don't know the possible mechanisms. Hence, we finally suggest that the geomagnetic jerks may be one kind of the excitation sources of the SYO and EYO.

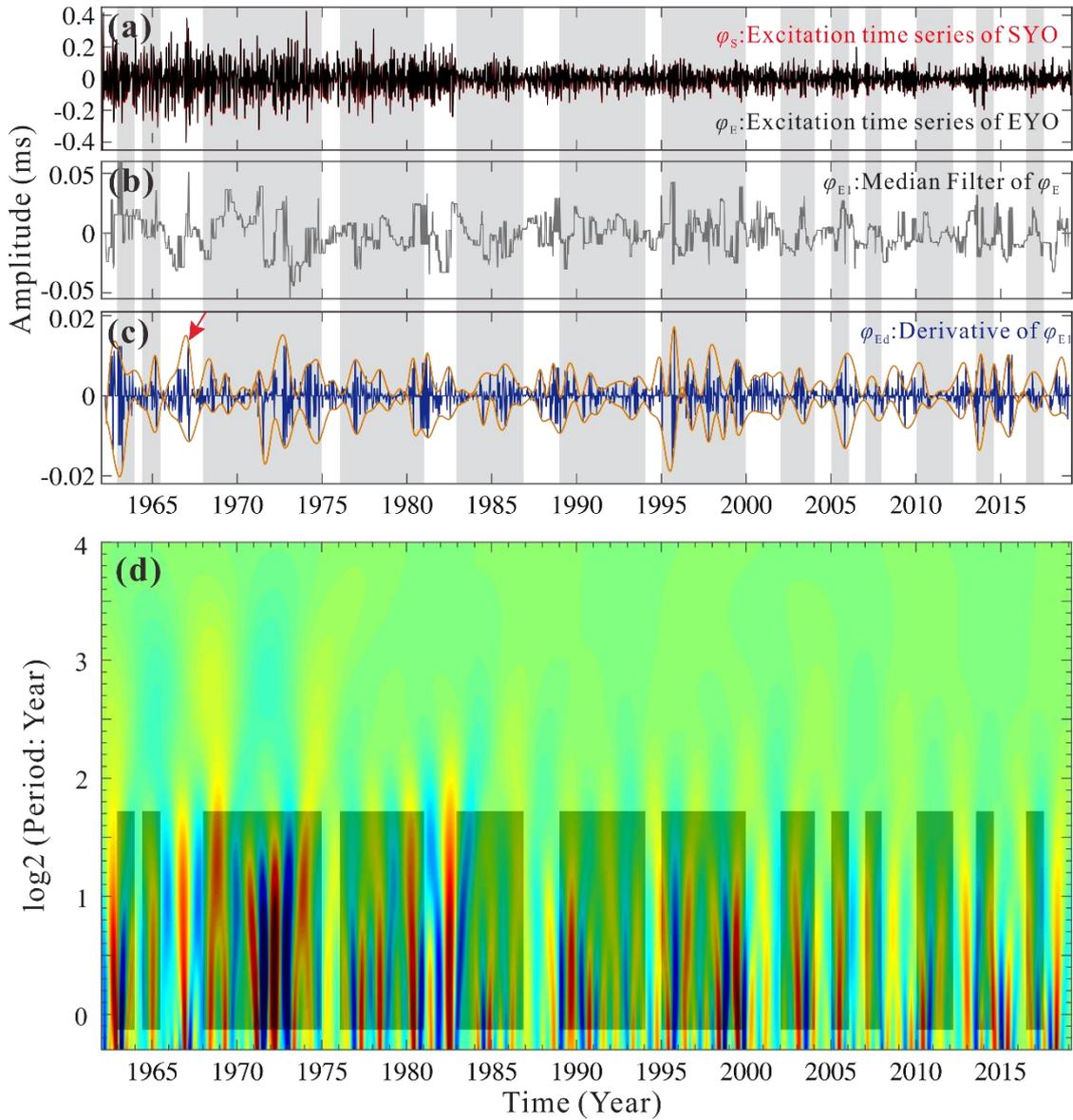

**Figure 9.** (a) The excitation functions $\varphi_S(t)$ and $\varphi_E(t)$ for the SYO and EYO, respectively; (b) $\varphi_{E1}(t)$: the filtered $\varphi_E(t)$ after using a median filter; (c) the first derivative of the $\varphi_{E1}(t)$; (d) the NMWT spectrum of the time series in (c). The gray areas in (a) and (b) indicate the jerks and jerks bounds from De Michelis et al. (1998), Brown, et al. (2013), Soloviev et al. (2017) and Hammer (2018).

## 5. Discussions and conclusions

The fluctuation characteristics of the SYO and EYO will affect the understanding of their physical mechanisms. To confirm these fluctuation characteristics, we reanalyzed the used ΔLOD records in used previous studies and further analyzed the EOPC04 ΔLOD time series based on the classic filter process. Our results show that the results for the SYO from the classic filter process are well consistent with the corresponding results in Holme and de Viron (2013) and Ding (2019) (both studies used a fitting and removing process); meanwhile, the spectra results reveal that the stable damping SYO and EYO suggested by DH20 are not consistent with the ΔLOD record that they used. Although there is a residual peak around ~7yr period in spectrum of $R_0–(S_1+E_1)$ (in Fig.2d), we confirm that it was just caused by the residual energy of the ~5.9yr and ~8.6yr signals (see Figs. 3 and 4). Our results from the classic filter process based on the EOPC04 and a much longer ΔLOD time series also show that the amplitudes of the SYO and EYO do not have stable damping trends in the 1962-2019 time span. Meanwhile, we find that an ~7.6yr signal is also present in the ΔLOD time series. Although Hao Ding has fitted and removed a 7.7yr signal in D19 to obtain a cleaner SYO time series, given that the amplitude of this signal is only about 0.025ms (almost the noise level), D19 didn't specifically explain it. In addition, because the mean amplitudes of the residual SYO and EYO in the results of DH20 are about 0.042ms, this ~7.6yr signal is also impossible can be found in spectra of DH20. To further confirm whether this ~7.6yr signal is a periodic signal, we use the OSE method to stack radial components of 35 Earth surface geomagnetic observations, the results show that there is a ~7.6yr periodic signal with $Y_{2,-2}$ spherical harmonic spatial distribution, and this signal has good phase consistency with the ~7.6yr signal in ΔLOD. In light of this, we tend to believe that the ~7.6yr signal in ΔLOD is a periodic signal.

For the possible relationship between jerks and SYO/EYO, DH20 have suggested that the jerks are well consistent with the peaks/valleys of the EYO, but they suggested that this finding is not hold for the SYO. We repeat the similar compassions as done by DH20. Different with DH20, our results show that only the 2017 jerk is well corresponding to the peak of the EYO, while there have 7 jerks which are well corresponding to the peaks/valleys of the SYO (see Fig. 7). Besides, considering that the jerk duration time even may be up to 4.7yr (Pinheiro et al. 2019), our results also indicate that the EYO cannot be used to predict jerks as suggested by DH20. We further compare the jerks with the sudden changes in the SYO/EYO time series (similar as done by Holme and de Viron (2013)) and their excitation time series (similar as done by D19), the results do demonstrate that the jerks seem to be related to the excitations of the SYO/EYO (see Figs. 7 and 9). Although we have suggested such possibility, considering the findings in Cox et al. (2016), further studies are still certainly needed.

Recalling the two disputes we summarized in section 1, in this study, we obtain the following conclusions:

1) We confirmed that the SYO and EYO have not exhibited stable damping trends since 1962. Instead, both of them have time-varying amplitudes. NMWT+BEPME methods used in previous studies may cause a strange increasing or decreasing trend for a stable cosine signal when there are many signals contained in the used time series.

2) Although the geomagnetic jerks do not have good consistency with the peaks/valleys of the SYO and EYO, the jerks do have some consistency with the sudden changes in the SYO/EYO time series and their corresponding excitation sequences. Such findings mean that the jerks are possible excitation sources for the SYO and EYO.

In addition, based on a deconvolution process, we estimate that the period $P$ and $Q$ values for the SYO and EYO are respectively [$P$=5.85±0.06yr, $Q$≥180] and [$P$=8.45±0.17yr, $Q$≥350]. The estimates for the SYO are very close to the values given in D19. We also add some explanations about the convolution/deconvolution in the Supplementary Information (see Fig. S3), which can help other researchers to understand the convolution/deconvoluted for the SYO and EYO.

**Acknowledgements**

We also thank L.T. Liu, X.Q. Su for NMWT algorithm and codes. The used ΔLOD data can be freely downloaded from provided by IERS (www.iers.org/IERS/EN/DataProducts/EarthOrientationData/eop.html); The AAM, OAM and HAM datasets were downloaded from: www.iers.org/IERS/EN/DataProducts/GeophysicalFluidsData/geoFluids.html; The geomagnetic dataset were downloaded from: http://www.wdc.bgs.ac.uk/dataportal/. If the readers have no access to the required non-standard licenced Matlab toolboxes, please contact with the first author. This work is supported by NSFC (grant #41974022, 41774024) and by Educational Commission of Hubei Province of China (grant #2020CFA109).

**References**

Abarca del Rio, R., Gambis, D. & Salstein, D. A. (2000). Interannual signals in length of day and atmospheric angular momentum. *Annales Geophysicae, 18*(3), 347-364. Doi: 10.1007/s00585-000-0347-9.

Alexandrescu, M., Gibert, D., Hulot, G., Le Mouël, J.L., & Saracco, G. (1995). Detection of geomagnetic jerks using wavelet analysis. *Journal of Geophysical Research: Solid Earth*, 100 (B7), 12557–12572.

Alexandrescu, M., Gibert, D., Hulot, G., Le Mouël, J.L., & Saracco, G. (1996). Worldwide wavelet analysis of geomagnetic jerks. *Journal of Geophysical Research: Solid Earth*, 101 (B10), 21975–21994.

Brown, W. J., Mound, J. E., & Livermore, P. W. (2013). Jerks abound: an analysis of geo-magnetic observatory data from 1957 to 2008. *Physics of the Earth & Planetary Interiors, 223*, 62-76.

Chao, B. F., Chung, W. Y., Shih, Z. R., & Hsieh, Y. K. (2014). Earth's rotation variations: a wavelet analysis. *Terra Nova, 26*(4), 260-264.

Chao, B.F., & Chung, W.Y. (2012). Amplitude and phase variations of Earth's Chandler wobble under continual excitation. *Journal of Geodynamics. 62*, 35–39

Chao, B. F., & Hsieh, Y. K. (2015). The Earth's free core nutation: formulation of dynamics and estimation of eigenperiod from the very-long-baseline interferometry data. *Earth and Planetary Science Letters*, *432*, 483–492.

Chao, B. F. (2017a), Dynamics of axial torsional libration under the mantle-inner core gravitational interaction, *Journal of Geophysical Research: Solid Earth*, *122*, 560– 571, Doi: 10.1002/2016JB013515.

Chao, B. F. (2017b). On rotational normal modes of the earth: resonance, excitation, convolution, deconvolution and all that. *Geodesy and Geodynamics. 8*(6), 371–376.

Chao, B. F., Yu, Y., & Chung, C. H. (2020). Variation of Earth's oblateness J2 on interannual-to-decadal timescales. *Journal of Geophysical Research: Solid Earth*, *125*, e2020JB019421. https://doi.org/10.1029/2020JB019421

Chulliat, A., Alken, P., and Maus, S. (2015), Fast equatorial waves propagating at the top of the Earth's core. *Geophysical Research Letters*, *42*, 3321- 3329. Doi:10.1002/2015GL064067.

Chulliat, A., & Maus, S. (2014). Geomagnetic secular acceleration, jerks, and a localized standing wave at the core surface from 2000 to 2010. *Journal of Geophysical Research: Solid Earth, 119*(3), 1531-1543. https://doi.org/10.1002/2013JB010604

Cox, G.A., Livermore, P.W., & Mound, J.E. (2016). The observational signature of modelled torsional waves and comparison to geomagnetic jerks, *Physics of the Earth and Planetary Interiors*. *255*, 50–65。

De Michelis, P., Cafarella, L., & Meloni, A. (1998). Worldwide character of the 1991 geomagnetic jerk. *Geophysical Research Letters*, 25 (3), 377–380.

Dill, R. (2008). Hydrological model LSDM for operational Earth rotation and gravity field variations. *Scientific Technical Report. 35, STR08/09,* GFZ Potsdam, Germany, https://doi.org/10.2312/GFZ.b103-08095 (2008).

Ding, H. (2019). Attenuation and excitation of the ~6 year oscillation in the length-of-day variation. *Earth and Planetary Science Letters, 507*, 131-139.

Ding, H., & Chao, B. F. (2018a). A 6-year westward rotary motion in the earth: detection and possible MICG coupling mechanism. *Earth and Planetary Science Letters, 295*, 50-55.

Ding, H., & Chao, B. F. (2018b). Application of stabilized AR-*z* spectrum in harmonic analysis for geophysics. *Journal of Geophysical Research*: *Solid Earth*, *123*, 8249-8259. Doi: 10.1029/2018JB015890.

Ding, H., & Chao, B. F. (2017). Solid pole tide in global GPS and superconducting gravimeter observations: Signal retrieval and inference for mantle anelasticity. *Earth and Planetary Science Letters*, *459*, 244-251. Doi: 10.1016/j.epsl.2016.11.039

Ding, H., & Chao, B. F. (2015a). Data stacking methods for isolation of normal-mode singlets of Earth's free oscillation: Extensions, comparisons, and applications. *Journal of Geophysical Research*: *Solid Earth*, *120*(7), 5034-5050. Doi: 10.1002/2015JB012025

Ding, H., Chao, B. F. (2015b). Detecting harmonic signals in a noisy time-series: the *z*-domain Autoregressive (*AR-z*) spectrum, *Geophysical Journal International, 201*(3), 1287–1296.

Ding, H., & Shen, W. B. (2013). Search for the Slichter modes based on a new method: optimal sequence estimation. *Journal of Geophysical Research*: *Solid Earth*, *118*(9), 5018-5029. Doi: 10.1002/jgrb.50344.

Duan, P. S., & Huang, C. L. (2020a). Intradecadal variations in length of day and their correspondence with geomagnetic jerks. *Nature Communications, 11*(1), 2273. https://doi.org/10.1038/s41467-020-16109-8

Duan, P. S., & Huang, C. L. (2020b). On the mantle-inner core gravitational oscillation under the action of the electromagnetic coupling effects. *Journal of Geophysical Research: Solid Earth*, *125*, e2019JB018863. https://doi.org/10.1029/2019JB018863

Duan, P. S. Liu, G. Y., Liu, L. T., Hu, X. G., Hao, X. G., Huang, Y., Zhang, Z. M., & Wang, B. B. (2015). Recovery of the 6-year signal in length of day and its long-term decreasing trend. *Earth Planets & Space, 67*(1), 161.

Duan, P. S., Liu, G. Y., Hu, X. G., Sun, Y. F, & Li, H. L. (2017). Possible damping model of the 6 year oscillation signal in length of day. *Physics of the Earth & Planetary Interiors, 265*, 35-42.

Duan, P. S., Liu, G. Y., Hu, X. G., Zhao, J., & Huang, C. L. (2018). Mechanism of the interannual oscillation in length of day and its constraint on the electromagnetic coupling at the core-mantle boundary. *Earth and Planetary Science Letters, 482*, 245-252.

Furuya, M., & Chao, B. F. (1996). Estimation of period and *Q* of the Chandler wobble. *Geophysical Journal International*, *127*, 693–702.

Gillet, N., Huder, L., & Aubert, J. (2019). A reduced stochastic model of core surface dynamics based on geodynamo simulations, *Geophysical Journal International*, *219*(1), 522–539. https://doi.org/10.1093/gji/ggz313

Gillet, N., Jault, D., & Finlay, C. C. (2015). Planetary gyre, time-dependent eddies, torsional waves, and equatorial jets at the earth's core surface. *Journal of Geophysical Research Solid Earth, 120*(6), 3991-4013. https://doi.org/10.1002/2014JB011786

Gillet, N., Jault, D., Canet, E., & Fournier, A. (2010). Fast torsional waves and strong magnetic field within the earth's core. *Nature, 465*(7294), 74. https://doi.org/10.1038/nature09010

Gross, R. S. (2005). The observed period and *Q* of the Chandler wobble. In: Plag, H.P., Chao, B.F., Gross, R.S., van Dam, T. (Eds.), *Forcing of Polar Motion in the Chandler Frequency Band: A*

*Contribution to Understanding Interannual Climate Change*, Cahiers du Centre Européen de Géodynamique et de Séismologie, Lux-embourg, *24*, 31–37.

Gross, R. S. (2015). Earth Rotation Variations - Long Period. *Treatise on Geophysics (Second Edition), 3*, 215-261.

Hammer, M. D. Local Estimation of the Earth's Core Magnetic Field. Ph.D. thesis, Technical University of Denmark (DTU), Kgs, Lyngby (2018).

Holme, R., & de Viron, O. (2005). Geomagnetic jerks and a high-resolution length-of-day profile for core studies. *Geophysical Journal International, 160*, 435-439.

Holme, R., de Viron, O. (2013). Characterization and implications of intradecadal variations in length of day. *Nature, 499*(7457), 202-4. https://doi.org/10.1038/nature12282

Liao, D. C., & Greiner-Mai, H. (1999). A new ΔLOD series in monthly intervals (1892.0–1997.0) and its comparison with other geophysical results. *Journal of Geodesy, 73*(9), 466-477. https://doi.org/10.1007/PL00004002

Liu, L., Hsu, H., & Grafarend, E. W. (2007), Normal Morlet wavelet transform and its application to the Earth's polar motion, *Journal of Geophysical Research: Solid Earth*, 112, B08401, doi:10.1029/2006JB004895.

Mound, J. E., & Buffett, B. A. (2006). Detection of a gravitational oscillation in length-of-day. *Earth and Planetary Science Letters, 243*(3-4), 383-389.

Petit, G. & Luzum, B., 2010. IERS Conventions 2010, IERS Technical Note; 36, Frankfurt am Main: Verlag des Bundesamts für Kartographie und Geodäsie, ISBN 3-89888-989-6.

Pinheiro, K. J., Jackson, A., & Finlay, C. C. (2013). Measurements and uncertainties of the occurrence time of the 1969, 1978, 1991, and 1999 geomagnetic jerks. *Geochemistry Geophysics Geosystems, 12*(10). Doi: 10.1029/2011GC003706

Pinheiro, K. J., Amit, H., & Terra-Nova, F. (2019). Geomagnetic jerk features produced using synthetic core flow models. *Physics of the Earth and Planetary Interiors*, *291*, 35–53.

Silva, L., Jackson, L., & Mound, J. (2012). Assessing the importance and expression of the 6 year geomagnetic oscillation. *Journal of Geophysical Research: Solid Earth*, 117, B10101. Doi: 10.1029/2012JB009405

Soloviev, A., Chulliat, A., & Bogoutdinov, S. (2017). Detection of secular acceleration pulses from magnetic observatory data. *Physics of the Earth & Planetary Interiors*, *270*, 128–142.

Torta, J. M., Pavón-Carrasco, F. J., Marsal, S., & Finlay, C.C. (2015). Evidence for a new geomagnetic jerk in 2014. *Geophysical Research Letters*, *42*, 7933–7940. https://doi .org /10 .1002 /2015GL065501.